\documentclass{emulateapj}

\shorttitle{Merger in HCG 31} \shortauthors{Amram et al.}

\begin{document}


\title{The Compact Group of Galaxies HCG 31 is in an early phase of merging\thanks{Based on observations collected
at the European Southern Observatory and at the Gemini North
telescope}}


\author{P. Amram \altaffilmark{1}, C. Mendes  de Oliveira \altaffilmark{2,3},
H. Plana \altaffilmark{4}, C. Balkowski \altaffilmark{5}, O.
Hernandez \altaffilmark{1,6}, C. Carignan \altaffilmark{6}, E.S.
Cypriano \altaffilmark{7,8}, L. Sodr\'e Jr. \altaffilmark{2}, J.L.
Gach \altaffilmark{1}  and J. Boulesteix \altaffilmark{1} }

\altaffiltext{1}{LAM/OAMP, Marseille, France.
(Philippe.Amram@oamp.fr)}

\altaffiltext{2}{IAG, S\~ao Paulo, Brazil.
(oliveira@astro.iag.usp.br)}

\altaffiltext{3}{Ludwig-Maximilians-Universit\"at, M\"unchen,
Germany}

\altaffiltext{4}{Universidade Estadual de Santa Cruz, Ilh\'eus,
Brazil}

\altaffiltext{5}{GEPI, CNRS et Paris 7, Meudon, France}

\altaffiltext{6}{Universit\'e de Montr\'eal and LAE, Montr\'eal,
Qu\'ebec, Canada.}

\altaffiltext{7}{Laborat\'orio Nacional de Astrof\'isica, Itajub\'a,
Brazil}

\altaffiltext{8}{Southern Astrophys. Research Telescope, La
Serena, Chile}




\begin{abstract}
We have obtained high spectral resolution (R = 45900) Fabry-Perot
velocity maps of the Hickson Compact Group HCG 31 in order  to revisit
the important problem of the merger nature of the central object A+C and
to derive the internal kinematics of the candidate tidal dwarf galaxies
in this group. Our main findings are: (1) double kinematic components
are present throughout the main body of A+C, which strongly suggests that
this complex is an ongoing merger (2) regions $A2$
and E, to the east and south of complex A+C, present rotation patterns
with velocity amplitudes of $\sim 25~km~s^{-1}$ and they
counterrotate with respect to A+C, (3) region F, which was previously
thought to be the best example of a tidal dwarf galaxy in HCG 31, presents
no rotation and negligible internal velocity dispersion, as is also the
case for region $A1$. HCG 31 presents an undergoing merger in its center
(A+C) and it is likely that it has suffered additional perturbations
due to interactions with the nearby galaxies B, G and Q.



\end{abstract}


\keywords{ galaxies: individual (HCG 31) --- galaxies: kinematics
and dynamics --- galaxies: evolution --- galaxies: interactions
--- galaxies: formation --- instrumentation: interferometers}


\section{Introduction}

The spectacular Hickson Compact Group 31 (Hickson 1982) shows a
wide range of  indicators  of galaxy interaction and merging:
tidal tails, irregular morphology, complex kinematics, vigourous
star bursting (e.g. Rubin et al. 1990) and possible formation of
tidal dwarf galaxies (e.g. Hunberger et al. 1996). All the objects
belonging to the group are embedded in a common large  HI envelope
(Williams et al. 1991). The group is formed by members A+C, B, E,
F, G, Q (Rubin et al. 1990).


Two scenarios have been put forward to explain the nature of the
central system A+C: it is either two systems that are about to
merge (e.g. Rubin et al. 1990) or a single interacting galaxy
(Richer et al. 2003). In this Letter we use our new Fabry-Perot
maps and deep imaging from Gemini-N to revisit the important
problem of the merger nature of the central object of the group,
A+C, and, in addition, we investigate the internal kinematics of
the tidal dwarf galaxies, in an attempt to identify if they are
self gravitating objects or not. We adopt a distance of 54.8 Mpc,
from the redshift z=0.0137 (Hickson et al. 1992) and using
H$_{\circ}$=75 km~s$^{-1}$~Mpc$^{-1}$, hence 1" $\sim$ 0.27 kpc.

\section{Observations}

Observations were carried out with the Fabry-Perot instrument
CIGALE (Gach et al. 2002) attached to the ESO 3.6m telescope in
August 2000. Interferograms were obtained with a high order
(p=1938) Fabry-Perot scanning interferometer, giving a free
spectral range of $155~km~s^{-1}$ with a \textit{Finesse
\textrm{F}=24} leading to a spectral resolution of $R=45~900$. The
pixel size is 0.405 arcsec; the total exposure time was 72 min (6
cycles of 12 min each, 48 scanning steps per cycle) and the FWHM
of the interference filter centered around 6651 $\AA$ was 20
$\AA$. The velocity sampling was 3 $km$ $s^{-1}$ and the relative
velocity accuracy is $\sim1~km~s^{-1}$ over the whole field where
the S/N is greater than 3. Reduction of the data cubes were
performed using the CIGALE/ADHOCw software (Boulesteix, 2002). The
data reduction procedure has been described e.g. in
Blais-Ouellette et al. (1999) and Garrido et al. (2002). For the
adopted distance of the group, one pixel corresponds to $\sim$
0.11 kpc.

In August 2003, we obtained two images with GMOS at Gemini-N in g'
and r' with exposure times of 1200 and 900 seconds respectively.
These images have a pixel size of 0.14" and typical seeing of
0.75".

\section{Results}


A color map of the group is presented in Fig. 1a.
It shows a wealth of star forming regions and the large extent of
the optical diffuse light which envelopes the group. For the first
time, regions $A1$ and $A2$, to the east of complex A+C, are seen
in great detail and depth.


Fig. 1b--d show the H$\alpha$ monochromatic map of the group, the
velocity map and several zoom panels showing the typical velocity
profiles in selected regions of the group. The velocity field was
corrected from free spectral range ambiguity using previous
kinematic observations. At a first look, galaxies A and C appear
to be a single kinematic entity, as their velocity fields show no
discontinuity. Under this assumption, we can infer a rough mass
for the A+C complex, within a radius of $\sim$ 18 arcsec ($\sim$
4.9 kpc) of $M\sim~4.5~\times~10^9~M_{\odot}$. The following
parameters and assumptions were used for this mass determination:
a maximum rotational velocity of 70$\pm 10~km~s^{-1}$ (only from
the NW side of the system given that the SE side is too
disturbed), a kinematic inclination and position angle of 51$\pm
5$ and 130$\pm 3$ degrees respectively and the assumption that the
measured motions are due to disk rotation. Nevertheless, multiple
profiles (see Fig. 1d), evident almost everywhere in the main body
of A+C,
strongly suggests that A+C is not a single entity and therefore
this rough determination provides only an order of magnitude for
the mass.

Fig. 2 shows the velocity gradients of the objects $A1$, $A2$, E
and F, situated around the pair A+C (see Fig 1a) and which have
been thought to be candidate tidal  dwarf galaxies (Hunsberger et
al. 1996, Iglesias-P\'aramo \& V\'{i}lchez 2001 and Richer et al.
2003). The curves are not corrected for inclination and the
central position and velocity were chosen such that the curves
were as symmetric as possible, with both sides matching, when
possible. The object with the highest velocity gradient is E, with
ordered velocities which range from 3950 $km~s$$^{-1}$ to 4000
$km~s$$^{-1}$. Object $A2$ also shows ordered motion, with
velocities going from 4125 to 4175 $km~s$$^{-1}$. Surprisingly,
object F, thought to be the best tidal dwarf galaxy candidate in
the group, shows a completely flat rotation curve, as does also
the smaller object to the northeast of A+C, fragment $A1$. In
addition, $A1$ and F have low gaseous velocity dispersions. These
results will be discussed in the next section.

We obtained the map of the gaseous velocity dispersion at each
pixel of the image (Amram et al. 2004, in preparation) assuming
the profiles are well represented by a single gaussian. The value
for the velocity dispersion ranges from 10 to 30 $km~s$$^{-1}$
throughout the group. In particular, objects $A1$, $A2$, E and F
show typical velocity dispersions of 15 $km~s$$^{-1}$, which in
some isolated regions can reach up to 25 $km~s$$^{-1}$. The
highest values for the velocity dispersion lie in the overlapping
region between A and C, mainly due to double components. It is
noticeable that these highest values do not match the most intense
star forming regions everywhere in the galaxies but particularly
where disk A and C overlap, implying that the line broadening and
the multiple components are not directly linked to star formation
triggered by interaction with another galaxy but specifically by
the merging of A+C. Hickson \& Menon (1985) reached a similar
conclusion, analyzing a radio continuum map of the group. They
found that the 20 cm peak of emission comes from the overlapping
regions of A and C, indicating additional evidence of recent
excessive starburst activity in this region.


\section{Discussion}

\subsection {The ongoing merger A+C}

Richer et al. (2003) also presented Fabry-Perot velocity maps of
HCG 31. There is fairly good general agreement between their
velocity field and ours. The spectral resolution of our maps is
nevertheless six times higher (7900 vs 45900) and the detection
limit several magnitudes fainter. Several authors supported that A
and C are separate entities in an ongoing merging phase (e.g.
Vorontsov-Velyaminov and Arhipova, 1963; Rubin et al., 1990) while
Richer et al. (2003) supported the scenario that A+C is a single
interacting spiral galaxy.  The new piece of evidence in support
of the merging scenario reported in this paper is the presence of
double velocity components throughout the system A+C. In fact, our
higher resolution velocity field shows kinematic structures which
are not naturally explained by a single disk but by the merging of
two disks.

Although Sc galaxies are, in general, transparent objects (Bosma
1995), moderate amounts of molecular gas (as traced by the CO) and
cold dust may make some regions opaque. The CO emission is weak in
HCG 31 but the brightest CO peak occurs in the overlapping region
between galaxies A and C (Yun et al. 1997) where the broader
H$\alpha$ profiles are observed. If the CO belongs to the
foreground galaxy, multiple components are observed in regions
optically thick and then could not be observed if disk A and C are
two separate galaxies seen in projection, i.e. chance alignment.
Multiple gaseous components are observed in the same disk plane
when they are not in equilibrium (e.g. \"{O}stlin et al. 2001).
This occurs when two different entities merge, or when the
feedback gas due to star formation interacts with the ISM. We
observe in HCG 31 the signature of both mechanisms, the second one
being probably a consequence of the first one.

The general pattern of the velocity field of HCG31 A+C is somewhat
similar to that of NGC 4038/9 (the Antennae, Amram et al., 1992)
in which a continuity in the isovelocities between both galaxies
is also observed in the overlapping region. The merging stage of
the Antennae is slightly less advanced than that for A+C. In the
Antennae, the two galaxies are clearly separate entities and their
bodies, which are not yet overlapping, each display an increasing
velocity gradient, which is roughly parallel and run from the NE
to the SW (Amram et al. 1992).
It is likely that when the disks of NGC 4038 and of NGC 4039
overlap, the velocity field will also present total continuity, as
observed in A+C.

Galaxy C (Mrk 1089) has been classified as a double nucleus
Markarian galaxy, the two nuclei being separated by 3.4 arcsec
(Mazzarella \& Boroson, 1993) and it is difficult to explain the
existence of the two nuclei without invoking a merging
 scenario, as shown by
numerical simulations (e.g. Barnes \& Hernquist 1992).  To
reproduce the double line profile in NGC 4848, Vollmer et al
(2001) have used numerical simulations. They interpret them as the
consequence of infalling gas which collides with the ISM within
the galaxy. This gives rise to an enhanced star formation observed
in the H$\alpha$ and in the 20 cm continuum map.


HCG 31 is most probably a group in an early phase of merger,
growing through slow and continuous acquisition of galaxies from
the associated environment. Moreover, several evidences for
interaction  with the other galaxy components of the group, namely
B, Q and G (e.g. the group is completely embedded in a large HI
envelope showing a local maximum on G to the SW and another one
around Q to the NE) indicate that the complex A+C is
most probably accreting the
surrounding galaxies.

\subsection {Tidal Fragments}

Several papers in the past (Hunsberger et al. 1996, Johnson and
Conti 2000, Richer et al. 2003 and Iglesias-P\'aramo \&
V\'{i}lchez 2001) have mentionned the possibility that tidal dwarf
galaxies in HCG 31 were formed. The best candidates are objects E
and F, for which metallicities were measured and they were
determined to be similar to that of the complex A+C, suggesting a
tidal origin for these objects (Richer et al., 2003).

As shown in Fig. 2, we detect ordered motions only for objects
$A2$ and E and not for $A1$ and F. It might be suspected that the
internal velocity motions measured in $A2$ and E could be due to
streaming motions in incipient tidal tails in formation (see Fig.
1a).  This is, however, not the case because these objects are
counterrotating with respect to the main body of A+C. The
discontinuity of the isovelocities can  be clearly seen from Fig.
1c: the velocities go from high (northeast) to low (southwest) in
A+C, towards object E. Then, along the body of E they go in the
opposite sense. Similarly, for object $A2$, it presents
counterrotation with respect to its immediate neighbor to the
west: galaxy A. These objects may fall back onto their progenitor.
In fact, from their velocity differences with respect to the
A+C complex (+115 $km~s$$^{-1}$ and -60 $km~s$$^{-1}$
respectively) and from their relative projected distances (6.7 kpc
and 5.4 kpc respectively) and assuming a total mass for the A+C
complex of $M\sim~4.5~\times~10^9~M_{\odot}$, we could determine
that these two
objects will indeed, most probably, fall back onto A+C. The
same is true for object F, which although more distant from A+C,
has a very small radial velocity difference of $\sim$ 60
$km~s$$^{-1}$. We note, however, that given the fact that we measure
radial velocities (and not the velocity component in the plane of the sky),
the observed internal velocities of the tidal fragments are lower limits.

Region F, which was previously thought to be the best example of a
tidal dwarf galaxy in HCG 31, is indeed part of the main merger,
following the same kinematic pattern of  the parent galaxy and
presenting no rotation nor significant internal velocity
dispersion. Region F was found to have low or inexistent old
stellar population by Johnson and Conti (2000).  If there is no
old stellar population the velocity dispersion of the gas is
mainly indicating the dynamics of the cloud. The range of values
derived for the internal velocity dispersion of F, 15-25
$km~s$$^{-1}$, is too close to the natural turbulence of the gas
and/or the expanding velocity due to starburst winds.
The lack of rotation and the continuity of the kinematics between
the main body of the merger and object F suggest it is simply a
tidal debris, although, considering its projected distance from
A+C (16 kpc) and the large amount of fuel available in the whole
area of the merger (2.1 $\times~10^{10}~M_\odot$ of HI gas), it
could perhaps develop into a tidal dwarf galaxy in the future, by accretion of
infalling material.

  There is also the possibility that regions $A1$ and F present no
velocity gradient due to their rotation pattern be along the line
of sight. Although this could be a possibility for the smaller and
rounder region $A1$, it is less likely the case for region F,
given its elongated morphology.


\section{Conclusions}

Our two main results are:

1) We measure multiple kinematic components throughout the body of
A+C which we interpret as a strong indication that this complex is
an ongoing merger.
The double photometric nucleus has been identified
in several previous images of HCG 31 including the spectacular HST
image published by Johnson and Conti (2000) and in Fig. 1a. The
double kinematic component is shown here for the first time in
Fig. 1d.

2) F and $A1$
present flat
rotation velocity profiles and insignificant velocity dispersions.
In contrast, $A2$ and E are
structures counterrotating with respect to the parent A+C complex.

We conclude that
HCG 31 is a merging group which is probably going to soon end up
as a field elliptical galaxy.
%
It would be very valuable to compare our new, high spectral
resolution maps to simulations of groups to investigate which
interaction and merger parameters fit these data.

\acknowledgments

The authors thank Olivier Boissin for help during the observations,
Jorge Iglesias-P\'{a}ramo for kindly providing the calibrated H$\alpha$
image and to acknowledge financial support from the
French-Brazilian PICS program.  CMdO, LS and ESC would like to thank
the Brazilian PRONEX program, FAPESP, CNPq.  CMdO deeply acknowledges
the funding and hospitality of the MPE Institut in Garching,
where
this work was finalized.  CC and OH acknowledge support from FQRNT,
Qu\'ebec and NSERC, Canada.

\begin{figure*}


\caption{\textit{HCG 31.}  \textbf{Upper Left, Fig. 1a:} Composite g' and r'
 color image of the central region of the group.
\textbf{Upper Right, Fig. 1b:} net-H$\alpha$ map. The flux is
given in a logarithmic scale, in units of
$10^{-16}~erg~s^{-1}~arcsec^{-2}~cm^{-2}$, the calibration has
been done with the H$\alpha$ image from Iglesias-P\'aramo \&
V\'ilchez (2001). The four small open squares labeled 11, 12, 21,
22 are the windows within which the profiles shown in Fig. 1d have
been extracted. Their size are 8$\times$8 pixels.
($\sim$3.2"$\times$3.2" or $\sim$0.9 kpc$\times$0.9 kpc).
\textbf{Bottom Left, Fig. 1c:} H$\alpha$ velocity field; the scale
is labeled in $km~s^{-1}$ and the black isocontours are separated
by 25 $km~s^{-1}$. \textbf{Bottom Right, Fig. 1d:} Examples of
velocity profiles corresponding to the windows displayed on the
H$\alpha$ image (Fig. 1b). Each small box represents 2$\times$2
pixels ($\sim$0.8"$\times$0.8" or $\sim$0.22 kpc$\times$0.22 kpc)
on the sky. The origin of the X-axis corresponds to 6648 $\AA$
(3917 $km~s^{-1}$) and the amplitude of the \textit{x-axis}
represents the $\sim 3.5 \AA$ free spectral range of the
interferometer ($\sim 155~km~s^{-1}$). The profiles are plotted
without correction for possible free spectral range jump: velocity
differences between various components may be the value plotted in
the figure plus n $\times~155~km~s^{-1}$ (n being an integer,
positive or negative).  The intensities of the profiles have been
normalized to the brightest one. The color background pixels
correspond to the intensities displayed on the H$\alpha$ image.}
\end{figure*}

\begin{figure}
\resizebox{8cm}{!}{\includegraphics{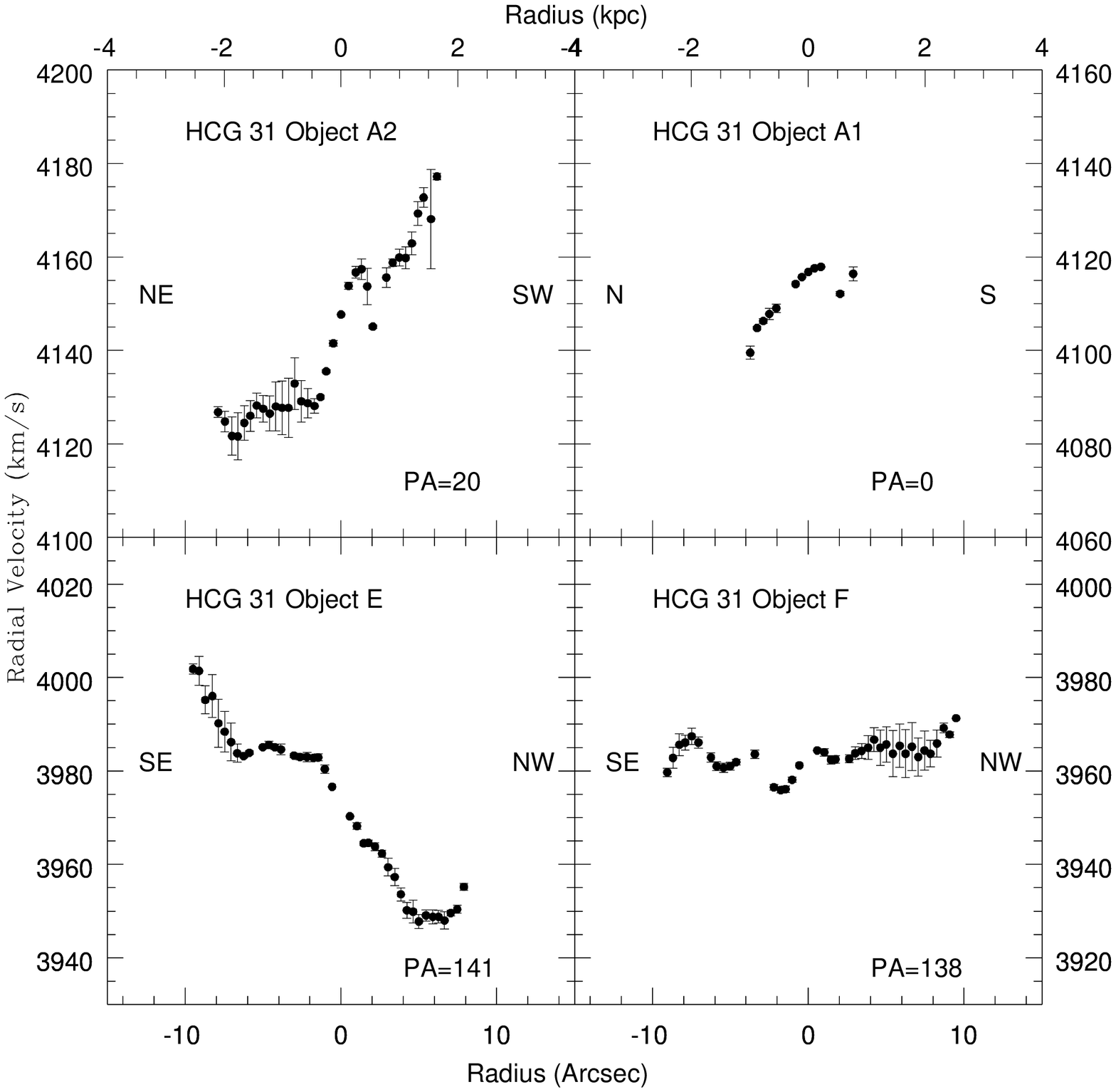}}
\caption{Velocity gradients for the tidal dwarf galaxy candidates}
\label{fig2}
\end{figure}

\clearpage






\end{document}